\documentclass[5p, final]{elsarticle}
\usepackage{geometry}
\usepackage{graphicx}
\usepackage{amssymb}
\usepackage{amsmath}
\usepackage[modulo]{lineno}
\usepackage{epstopdf}
\usepackage{algorithm}
\usepackage{algpseudocode}
\usepackage{verbatim}
\usepackage{url}
\usepackage{hyperref}
\usepackage{mathtools}
\DeclarePairedDelimiterX{\infdivx}[2]{(}{)}{%
  #1\;\delimsize\|\;#2%
}
\newcommand{\KL}{\text{KL}\infdivx}
\journal{NME}

\title{Developing Deep Learning Algorithms for Inferring Upstream Separatrix Density at JET}
\author[HEL]{A. Kit\corref{cor1}}
\ead{adam.kit@helsinki.fi}
\author[HEL,VTT]{A.E. Järvinen}
\author[FZJ]{S. Wiesen}
\author[EIN]{Y. Poels}
\author[KTH]{L. Frassinetti}
\author[JET]{JET Contributors}

\cortext[cor1]{Corresponding author: adam.kit@helsinki.fi}
\address[HEL]{University of Helsinki,FI-00014, Helsinki, Finland}
\address[VTT]{VTT Technical Research Centre of Finland, FI-02044 VTT, Finland}
\address[FZJ]{Forschungszentrum Jülich GmbH, Institut für Energie- und Klimaforschung– Plasmaphysik, DE-52425 Jülich, Germany}
\address[EIN]{Eindhoven University of Technology, NL-5600 MB, Eindhoven, Netherlands}
\address[KTH]{KTH Royal Institute of Technology, SE-1142B, Stockholm, Sweden}
\address[JET]{See the author list of ‘Overview of JET results for optimising ITER operation’ by J. Mailloux et al. to be published in Nuclear Fusion Special issue: Overview and Summary Papers from the 28th Fusion Energy Conference (Nice, France, 10-15 May 2021)}

\begin{document}

\begin{abstract}
    Predictive and real-time inference capability for the upstream separatrix electron density, $n_\text{e, sep}$, is essential for design and control of core-edge integrated plasma scenarios. In this study, both supervised and semi-supervised machine learning algorithms are explored to establish direct mapping as well as indirect compressed representation of the pedestal profiles for predictions and inference of $n_{\text{e, sep}}$. Based on the EUROfusion pedestal database for JET \cite{Frassinetti_NF_2021}, a tabular dataset was created, consisting of machine parameters, fraction of ELM cycle, high resolution Thomson scattering profiles of electron density and temperature, and $n_{\text{e, sep}}$ for 608 JET shots. Using the tabular dataset, the direct mapping approach provides a mapping of machine parameters and ELM percentage to $n_{\text{e, sep}}$. Through representation learning, a compressed representation of the experimental pedestal electron density and temperature profiles is established. By conditioning the representation with machine control parameters, a probabilistic generative predictive model is established. For prediction, the machine parameters can be used to establish a conditional distribution of the compressed pedestal profiles, and the decoder that is trained as part of the algorithm can be used to decode the compressed representation back to full pedestal profiles. Although, in this work, a proof-of-principle for predicting and inferring $n_{\text{e, sep}}$ is given, such a representation learning can be used also for many other applications as the full pedestal profile is predicted.   An implementation of this work can be found at \url{https://github.com/fusionby2030/psi_2022}. 
\end{abstract}

\maketitle

\section{Introduction}
The electron density at the upstream separatrix, $n_{\text{e, sep}}$, is a key parameter for core-edge integration in tokamaks. The power exhaust properties and scrape-off layer (SOL) are strongly regulated by $n_{\text{e, sep}}$ and collisionality, with power exhaust requirements typically favoring elevated $n_{\text{e, sep}}$ \cite{Stangeby}. 
On the other hand, elevated $n_{\text{e, sep}}/n_{\text{e, ped}}$ and $n_{sep}$/$n_\text{GW}$ ratios have been observed to lead to degraded core and pedestal performance in high confinement mode (H-mode) plasmas \cite{Frass_NF_21_nesep, Eich_NF_2018}. Therefore, predictive and real-time inference capability for $n_{\text{e, sep}}$ is essential for design and control of core-edge integrated plasma scenarios. 

Due to the multiple physical processes impacting $n_\text{e,sep}$, there are no first-principles predictive models that would predict $n_\text{e,sep}$ based on machine control parameters only. 2D SOL multi-fluid simulation packages, such as SOLPS-ITER \cite{Wiesen_NM_2015}, can in principle calculate a value for $n_\text{e,sep}$. However, the predicted value depends on user-specified phenomenological parameters, such radial diffusion and heat conduction coefficients, as well as assumptions on recycling and pumping. To address the gap of predictive capability for $n_\text{e,sep}$, empirical scaling laws have been derived based on H-mode databases \cite{Frassinetti_NF_2021, Kallenbach_2018}. The scaling law by Frassinetti et al. is based on the JET pedestal database (JET-PDB) \cite{Frassinetti_NF_2021}. The data is in tabular form with rows (entries) of shots that have columns (features) corresponding to machine control parameters and measurements of various plasma parameters averaged over a stationary time window. The plasma parameters and machine parameters come from a time window corresponding to a flat-top (normally around 1-2 seconds) of an H-mode pulse. The plasma parameters, $n_{\text{e, sep}}$ included, are determined from a modified hyperbolic tangent (mtanh) \cite{MTANH_Frass, MTANH_Groebner} fit of the electron temperature, $T_e$, and density, $n_e$, profiles from the High Resolution Thomson Scattering (HRTS) system at JET \cite{Frassinetti_NF_2021, Frass_NF_21_nesep}. The profiles used in the curve fitting are those that fall between 75-99\% of an Edge Localized Mode (ELM) cycle within the flat-top window. In this analysis, we use JET-PDB to go beyond log-linear regression using advanced machine learning (ML) tools, as well as use the pulses and time-windows given in the database to create a larger dataset that is used to fit ML models to predict $n_\text{e,sep}$, as well as full HRTS $n_e$ and $T_e$ profiles. 

In JET-PDB, $n_\text{e,sep}$ is determined from the fitted HRTS profiles by assuming a conduction limited SOL and calculating the upstream $T_e$ based on Spitzer-Härm heat conductivity.  This can be considered a two-point model approach for determining the separatrix location \cite{Stangeby}. The method of determining $n_\text{e,sep}$ from HRTS profiles is outlined in detail in \cite{Stangeby_nesep, Leonard_nesep,Frassinetti_NF_2021}, but relevant steps are restated here. As HRTS measures $n_e$ and $T_e$ for the same spatial location, power balance arguments can be used to align $T_e$ with separatrix. Due to the strong temperature dependence of Spitzer-Härm heat conductivity $\propto T_e^{5/2}$, the upstream $T_e$ in conduction limited SOL scales strongly sub-linearly as a function of heat-flux in the SOL, $T_e \propto q_{||}^{2/7}$. As a result, for typical JET H-mode conditions, $T_e$ can be approximated to be around 100 eV at the upstream separatrix.  Fitting the profiles that fall between 75-99\% of the ELM cycles within the given flat-top H-mode time window, the value of the fitted $T_e$ closest to 100 eV is, therefore, the location of the corresponding density point at the separatrix, $n_\text{e,sep}$. The corresponding machine parameters for this entry are considered to be the average across the entire time window. 
\begin{figure}
    \centering
    \includegraphics[scale=0.38]{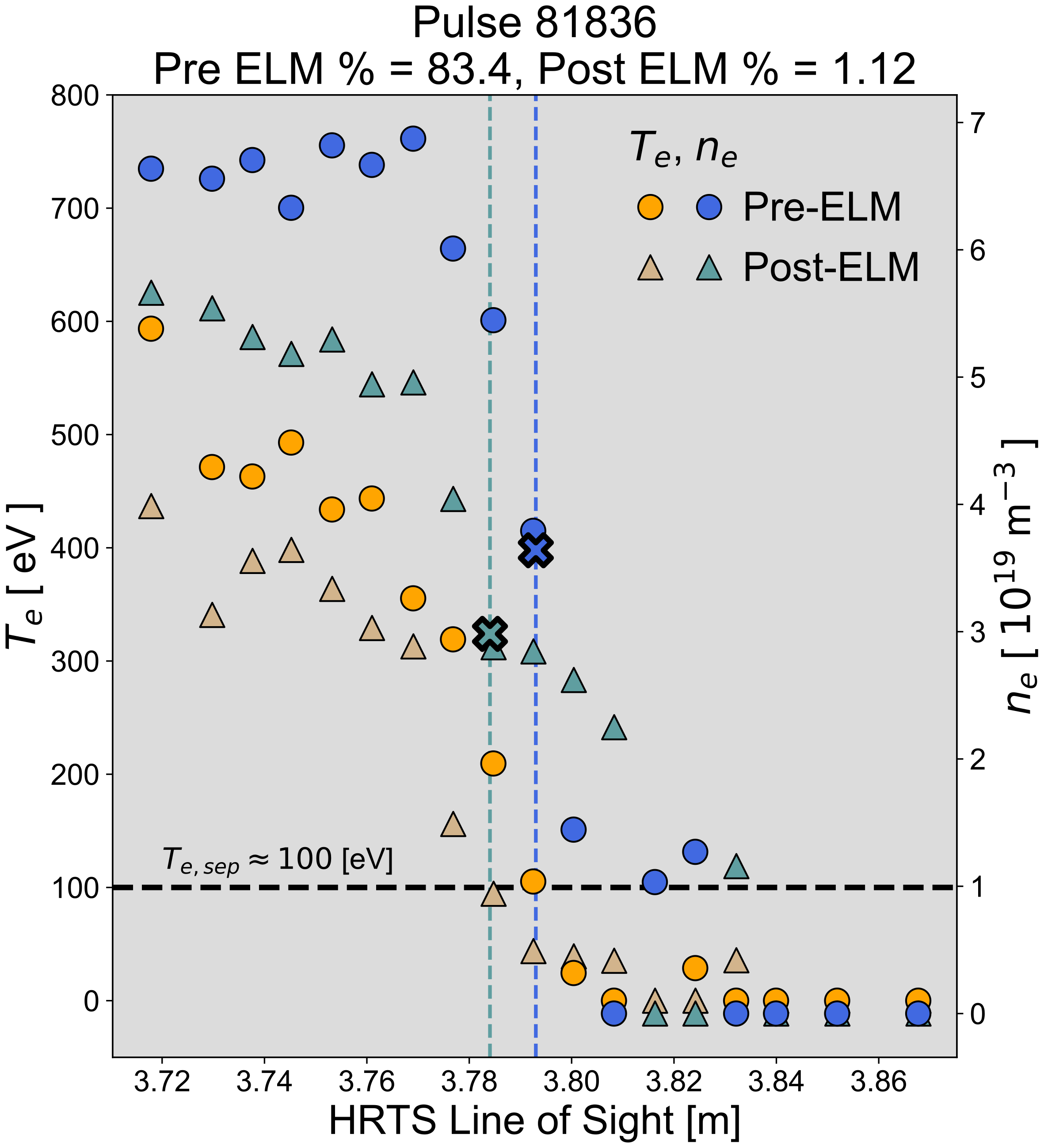}
    \caption{ 
    An example of the $n_{\text{e, sep}}$ estimations (solid colored X) for JET pulse 81836 for varying times within an ELM cycle. The characteristic crashing of density gradients in ELMy H-modes will lead to variations in $n_{\text{e, sep}}$. The assumption of $T_e \sim 100$ eV is not expected to be valid throughout the ELM cycle due to the fluctuations of the power crossing the separatrix.Also in pulses with very high ELM frequency, it is possible that the linear approximation method breaks due to the rapid fluctuation of the measured plasma quantities, relative to the HRTS measurement frequency. These shortcomings are acknowledged here and will be taken into account in future studies.}
    \label{fig:FIND_NESEP}
\end{figure}

\section{Dataset}
In this analysis we avoid mtanh curve fitting and ELM filtering and instead calculate $n_\text{e,sep}$ via a linear approximation of the HRTS measurements, which increases the amount of data points, w.r.t the JET-PDB,  by an order of magnitude. This approach neglects the impact of the non-zero instrument function that would be taken into account in the mtanh fitting procedure, but the impact of this correction is a second order effect compared to shot to shot variation due to different engineering parameters \cite{MTANH_Frass}.
 To estimate $n_\text{e,sep}$ for a given time slice without time/elm-averaging, we apply the following linear approximation method (Fig \ref{fig:FIND_NESEP}):

\begin{enumerate}
    \item Make an initial guess of where the separatrix is located $r_{\text{sep}} \approx \frac{1}{2} \left( r_{\text{top}} + r_{\text{bot}} \right)$, where $r_{\text{top}}, r_{\text{bot}}$ are pedestal top and bottom determined from the second derivative of density profile, respectively.
    \item Find the closest two temperature points, $T_L, T_R$, to the initial $r_{\text{sep}}$, s.t., $T_L < 100 \text{ eV} < T_R$. 
    \item Find weights, $w_L, w_R$, via linear approximation, s.t., $w_LT_L + w_RT_r = 100 \text{ eV}$ and $w_L + w_R = 1$. 
    \item Use weights: $w_Ln_L + w_Rn_R = n_{\text{e, sep}}$ and $w_Lr_L + w_Rr_R = r_{\text{sep}}$, where $n_L, n_R, T_L, T_R$ are the electron densities and temperatures located at radial positions $r_L, r_R$.   
\end{enumerate}  
 This approximation relies heavily on the assumption that $T_\text{e, sep} = 100$ eV, which is often not the case during the course of an ELM cycle. Future studies will address this issue by applying other models for $T_\text{e, sep}$. A possible pathway is to apply SOLPS-ITER, or a fast ML surrogate model for SOLPS-ITER \cite{Dasbach_this_conference}, to provide a more sophisticated prediction of $T_\text{e, sep}$ than is obtained by the simple two-point model. Additionally, $n_{\text{e, sep}}$ values determined from the linear approximation are typically smaller than those found in the JET-PDB, as in the JET-PDB only profiles from 75-99\% of the ELM cycle are used. For now, we use this linear approximation to estimate $n_{\text{e, sep}}$ for a given profile, and move to constructing a dataset to be able to map machine parameters to each profile/$n_{\text{e, sep}}$. 

A dataset was created, including machine parameters, $n_{\text{e, sep}}$, and HRTS measurements in the edge, using a subset of pulses from the JET-PDB \cite{Frassinetti_NF_2021}. The following criteria were applied: JET-ILW, deuterium fuelling, no seeding or nitrogen seeding, and no ELM mitigation techniques (kicks, pellets, or resonant magnetic perturbations) applied. This subset consists of over 600 shots, with a total of around 20000 time slices. Each entry in the dataset consists of time slice of HRTS measurements with corresponding fraction of ELM cycle. For a given time slice, the machine parameters are the average of those that fall between a window of 20 ms before the HRTS measurement (the HRTS measurement frequency); an overview of the domains is given in Table  \ref{tab:domains}. For each time slice, only the last 20 points of the HRTS profile is used, from which $n_{\text{e, sep}}$ is calculated via the linear approximation. Thus, we expand the amount of entries that could be used to fit a regression-like model by an order of magnitude (600 $\rightarrow$ 20000). Armed with the larger dataset, we set about predicting $n_{\text{e, sep}}$. 

\begin{table}
    \caption{Machine parameters and their domains for the constructed dataset. Unitless parameters are denoted as [-]. All variables except the gas puff and power parameters are taken from the equilibrium reconstruction for each pulse. The ELM fraction is the percentage of the ELM cycle that the HRTS measurement was taken. $q_{\text{cycl}}$ is the cylindrical approximation for $q_{95}$, since it could be argued that $q_{95}$ would contain information regarding the separatrix.
    We calculate $q_{\text{cycl}}$ using a cylindrical approximation for the tokamak cross section, parameterized by relevant machine parameters, therefore $q_{\text{cycl}} : q_{\text{cycl}}(\kappa, a, \delta, B_T, R_0, I_P)$. Although this information is essentially encoded by other parameters, this can be seen as a form of feature engineering.} 
    \centering
    \begin{tabular}{|c|c||c|c|}
        \hline 
        Param [Unit] & Min, Max & Param [Unit] & Min, Max \\
        \hline 
        $R_0$ [m]& 2.81, 2.97 & $P_{\text{NBI}}$ [MW] & 0.9, 32 \\
        $a$ [m]& 0.87, 0.96 & $P_{\text{ICRH}}$ [MW] & 0.0, 7.3 \\ 
        $\delta_U$ [-] & 0.08, 0.47 & $P_{\text{OHM}}$ [MW] & 0.1, 2.27\\ 
        $\delta_L$ [-] & 0.23, 0.49 & $B_T$ [T] & 0.96, 3.7  \\ 
        $\kappa$ [-] & 1.49, 1.83 & $I_P$ [MA] & 0.96, 3.98 \\ 
        $V_P$ [m$^{3}$] & 69, 80 & $\Gamma$ [e$^{23}$/s] & 0, 1.3 \\
        $q_{\text{approx}}$ [-] & 2.41, 5.49 & ELM Frac. [-] & 1, 99 \\ 
        \hline 
    \end{tabular}
    
    \label{tab:domains}
\end{table}

\section{Machine Learning for Predictive Modeling}
To predict $n_{\text{e, sep}}$ from machine parameters, we propose two approaches; direct mapping and representation learning. The direct mapping approach aims to learn a mapping from machine parameters \textbf{directly} to $n_{\text{e, sep}}$. The representation learning approach aims to learn a lower dimensional \textbf{representation}, denoted as $z$, of the HRTS profiles, as opposed to only predicting $n_{\text{e, sep}}$ as per the direct mapping. The lower dimensional representation is conditioned on machine parameters and can be decoded to generate profiles. From the generated profiles, $n_{\text{e, sep}}$ is determined. Both approaches leverage recent advances in machine learning and deep neural networks. 

The direct mapping approach aims to learn a function $f: X \rightarrow n_{\text{e, sep}}$, where $X$ are the given set of machine parameters that lead to a particular $n_{\text{e, sep}}$. No profile data is given in this approach. We aim to learn $f$ by using the created dataset via supervised learning. Regression based supervised learning has seen major breakthroughs via artificial neural networks and decision tree ensembles in the last decades. We choose a decision tree ensemble, namely XGBoost \cite{XGBOOST}, to learn the direct mapping. XGBoost has shown good performance across many different types of tabular datasets \cite{TABOVER}. The learning goal for the direct mapping approach boils down to minimizing the mean-squared-error (MSE) of predicted $n_{\text{e, sep}}$ against the `true' $n_{\text{e, sep}}$ for a given set of machine parameters. 

\begin{figure*}
    \centering
    \includegraphics[scale=2.0]{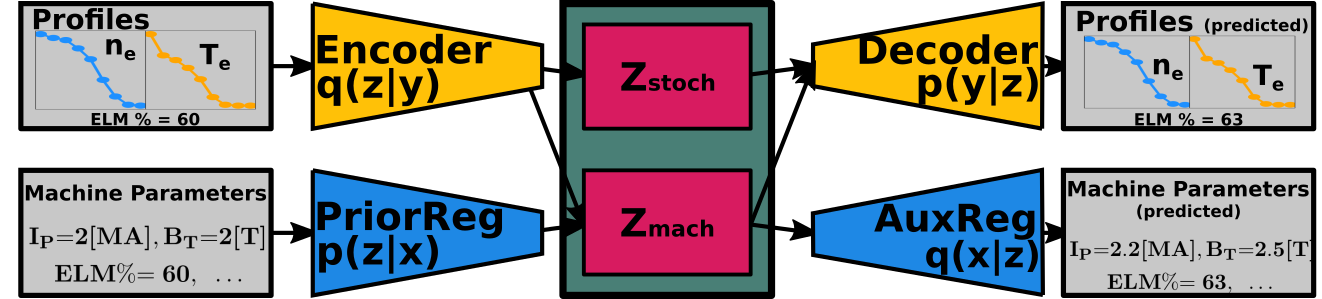}
    \caption{The architecture of the model used in the representation learning approach. This is a modification of the DIVA  framework \cite{DIVA}. For conditional generation, only the PriorReg and Decoder are used, in which machine parameters are passed through the PriorReg, from which latent variables $z$ are determined, which are then passed to the decoder, which produces density and temperature profiles.}
    \label{fig:VAE_ARCH}
\end{figure*}

The representation learning approach aims to learn a latent variable $z$, with prior distribution $p(z) = \mathcal{N}(0, 1)$, that represents profiles $y$ via $p(y, z) = p(y\mid z)p(z)$. This idea is implemented with the variational autoencoder (VAE) \cite{VAE} framework (Fig. \ref{fig:VAE_ARCH}), using an encoder $q(z\mid y)$, and decoder $p(y\mid z)$ distribution, parameterized by artificial neural networks. Contextually, this approach would take profiles as inputs, encode the information to a lower dimensional latent variable, then decode from the latent variable to profiles again. As the true likelihood of the data $p(y, z)$ is intractable, we optimize a lower bound, the Evidence Lower Bound (ELBO). For our chosen prior, maximizing the ELBO corresponds to minimizing the following function: 
\begin{equation*}\label{eq:VAE}
    \mathcal{L}_{\text{VAE}} \equiv \text{MSE}(y, \hat{y}) \\
    + \KL{q(z\mid y)}{\mathcal{N}(0, 1)}
\end{equation*}
i.e., the model learns to minimize the mean-squared-error (MSE) of predicted profiles $\hat{y}$ via the lower-dimension representation $q(z\mid y)$ determined from an encoder, against the `real' profile $y$, passed to the encoder. The latent representation $q(z\mid y)$ is regularized via the Kullback–Leibler divergence (KL) relative to a standard normal distribution, optimizing the latent distribution to closely follow a normal distribution with mean of 0 and variance of 1. 

We add an additional learning goal of predicting machine parameters from the latent variable $z$. This is done via an auxiliary regressor, which takes the encoded representation, $z$, and predicts machine parameters, i.e., $p(x\mid z)$. Ideally, this enforces the compressed representation to be informed about the machine configuration, i.e., the latent representation should not only encode profile information, but also encode the corresponding machine parameters for the given profile. The learning goal is updated to reflect this: 
\begin{equation*}\label{eq:part_two}
    \mathcal{L} \equiv \text{MSE}(y, \hat{y}) + \text{MSE}(x, \hat{x})  + \KL{q(z\mid y)}{ \mathcal{N}(0, 1)}
\end{equation*}
where the MSE between predicted machine parameters $\hat{x}$ and real machine parameters, $x$, is added. 

As per the DIVA framework \cite{DIVA}, we split the latent variable $z$ into different sub-spaces, namely, $z_{\text{mach}}$ and $z_{\text{stoch}}$. The auxiliarly regressor previously mentioned then only uses the latent variables from $z_{\text{mach}}$, and is therefore parameterized as $p(x\mid z_{\text{mach}})$. Ideally, all information pertaining to machine parameters would be contained in the latent variable $z_{\text{mach}}$ and information that represents the stochastic nature of the data would be contained in $z_{\text{stoch}}$. The learning goal can be updated to reflect this: 
\begin{equation*}\label{eq:part_three}
\begin{split}
    \mathcal{L} \equiv \text{MSE}(y, \hat{y}) + \text{MSE}(x, \hat{x})  \\
    + \KL{q(z_{\text{mach}}\mid y)}{\mathcal{N}(0, 1)} \\
    + \KL{q(z_{\text{stoch}}\mid y)}{\mathcal{N}(0, 1)}  \\
\end{split}
\end{equation*}

In order to learn a mapping of machine parameters to profiles, we introduce a conditional prior, $x:p(z_{\text{mach}}\mid x)$, to go from machine parameters to $z$. As per the DIVA framework, we want to learn a latent variable that is split by domain by using conditional priors depending on the domain of the data (machine parameters) during optimization. To learn this split we use conditional priors and the KL divergence. The conditional prior is the initial distribution found through machine parameters, then we optimize the approximate posterior ($z$ found through the regular encoder) to be as close to the prior via the KL divergence. Thus, via the conditional prior, we can pass machine parameters to generate profiles for prediction. The resulting learning goal for the representation learning approach is to minimize via the following function:
\begin{flalign} \label{eq:loss_og}
    \mathcal{L}_{\text{DIVA}} & \equiv \text{MSE}(y, \hat{y}) + \text{MSE}(x, \hat{x}) \\ 
    & + \KL{q(z_{\text{mach}}\mid y)}{p(z_{\text{mach}}\mid x)} \\ 
    & + \KL{q(z_{\text{stoch}}\mid y)}{\mathcal{N}(0, 1)}
\end{flalign}
Row 1 enforces the model to learn a representation to best `reconstruct' profiles and machine parameters. Row 2 enforces that the priors determined via the profile encoding are as close as possible to those suggested via the conditional prior found from machine parameters. Row 3 enforces the stochastic split latent variable, $z_{\text{stoch}}$, to be close to a standard normal distribution. 

We further impose physics constraints calculated on the generated profiles and machine parameters in order to enforce the model to learn to generate `physically realistic' or consistent profiles and machine parameters. This is done via the following three approximations: 
\begin{itemize}
    \item static stored pressure at the pedestal/SOL, $W_{e}
    ^\text{ped} := W_{e}
    ^\text{ped}(n_e, T_e) = k_B \sum_{k=0}^n p_{e, k}$ for $k$ radial measurements along the profile, and $p_e = n_eT_e$, the plasma pressure. This constraint uses the outputs of the decoder, i.e, the predicted profiles, and ideally would enforce the model to learn that pressure is a conserved quantity, i.e., if density increases,  temperature decreases. 
    \item poloidal magnetic field, \\ $B_{\theta}^\text{ped} := B_{\theta}^\text{ped}(I_P, \kappa, a, \delta_U, \delta_L) = \frac{\mu_0 I_P}{C}$, where $C$ is the approximation of circumference of a D-shaped cross section of JET. This constraint uses the outputs of the auxiliary regressor, i.e., the predicted machine parameters, and ideally would enforce the model to learn to scale the shaping parameters w.r.t to current, i.e., if one of these quantity changes and $B_{\theta, \text{ped}}$ would like to be conserved, then the other parameters need to be scaled relative to the changes. 
    \item ratio of magnetic pressure to plasma pressure $\beta_{e}^\text{ped} := \beta_{e}^\text{ped}(n_e, T_e, B_{\theta}^\text{ped}, B_T) = \frac{4 \mu_0 p_{e, 0}}{B_T^2 + B_{\theta}^{\text{ped}2}}$, where $p_{e, 0}$ is the plasma electron pressure measurement closest to the high field side. $B_{\theta}^\text{ped }$ is the same as calculated above.  This constraint uses both predicted machine parameters and profiles from the model, and ideally would enforce the model to learn to encode the scalings of parameters similarly to suggested above, i.e., if a plasma with different magnetic field or current has the same$\beta_{e}^\text{ped}$, then other parameters need to change to reflect this conserved quantity. 
\end{itemize}   

The learning goal then gains the following additional mean-squared-error terms:  
\begin{flalign*}
    \mathcal{L} \equiv &  \mathcal{L}_{\text{DIVA}} + \text{MSE}(W_{e}^\text{ped}, \hat{W}_{e}^\text{ped}) \\ 
    & + \text{MSE}(B_{\theta}^\text{ped}, \hat{B}_{\theta}^\text{ped})  + \text{MSE}(\beta_{e}^\text{ped}, \hat{\beta}_{e}^\text{ped})
\end{flalign*}
Ideally this enforces the model to learn a representation that is physically consistent.  There are many other physics based constraints that could be introduced, but these three show the proof-of-principle on how to introduce physics into the model directly via the learning goal. 

The benefit of representation learning is that it learns a physically interpretative lower dimensional representation of the plasma state that can be sampled for probabilistic predictions. 

\textbf{Physical interpretation}: Full density and temperature profiles are generated for each prediction, unlike the direct mapping. This allows us to validate the output of the model in terms of physical quantities. For example, we can inspect the conservation of the electron pressure ($p_e = n_eT_e$), i.e., if density increases, temperature should decrease. Another example is to inspect the inferred relationship between current, $I_P$, and density (Fig. \ref{fig:current}). 

\begin{figure*}
    \centering
    \includegraphics[width=0.8\textwidth]{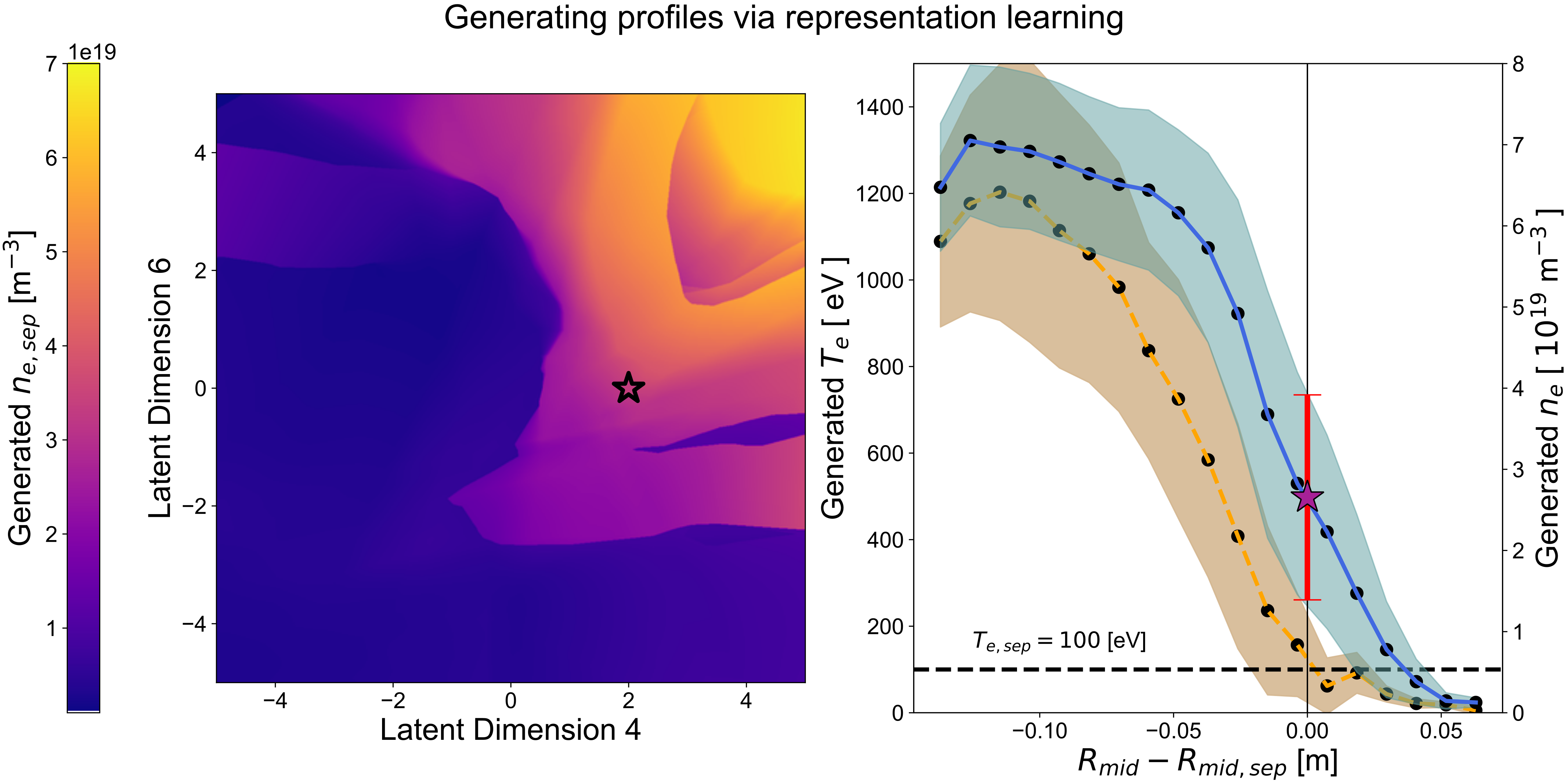}
    \caption{The low-dimensional latent variable $z$ is the learned representation of the plasma state. To investigate $z$, dimensions  4 and 6 are traversed while keeping remaining dimensions fixed. The relationship of this traversal is visualized with the inferred $n_{\text{e, sep}}$, high and low values of $n_{\text{e, sep}}$. Dimensions 4 and 6 were chosen because they had a high linear correlation with inferred $n_{\text{e, sep}}$.}
    \label{fig:HEADLINER}
\end{figure*}

\textbf{Latent representation}: Through the lower dimensional compressed representation we can identify nonlinear relations between the plasma state and machine parameters. For example, we could correlate latent dimensions to machine parameters and potentially find a set of machine parameters that leads to a specific $n_{\text{e, sep}}$ (Fig. \ref{fig:HEADLINER}). This has enormous potential, as the representation could be used as a basis for a control-like algorithm.

\textbf{Probabilistic}: We can quantify the uncertainty in the profile predictions as well as the machine parameters through latent variable $z_{\text{stoch}}$. This is done by repeatedly sampling from $z_{\text{stoch}}$, and taking the standard deviation across the samples (Fig. \ref{fig:HEADLINER}).

\section{Results}

\begin{figure*}
    \centering
    \begin{minipage}[b]{0.45\textwidth}
    \includegraphics[width=\textwidth]{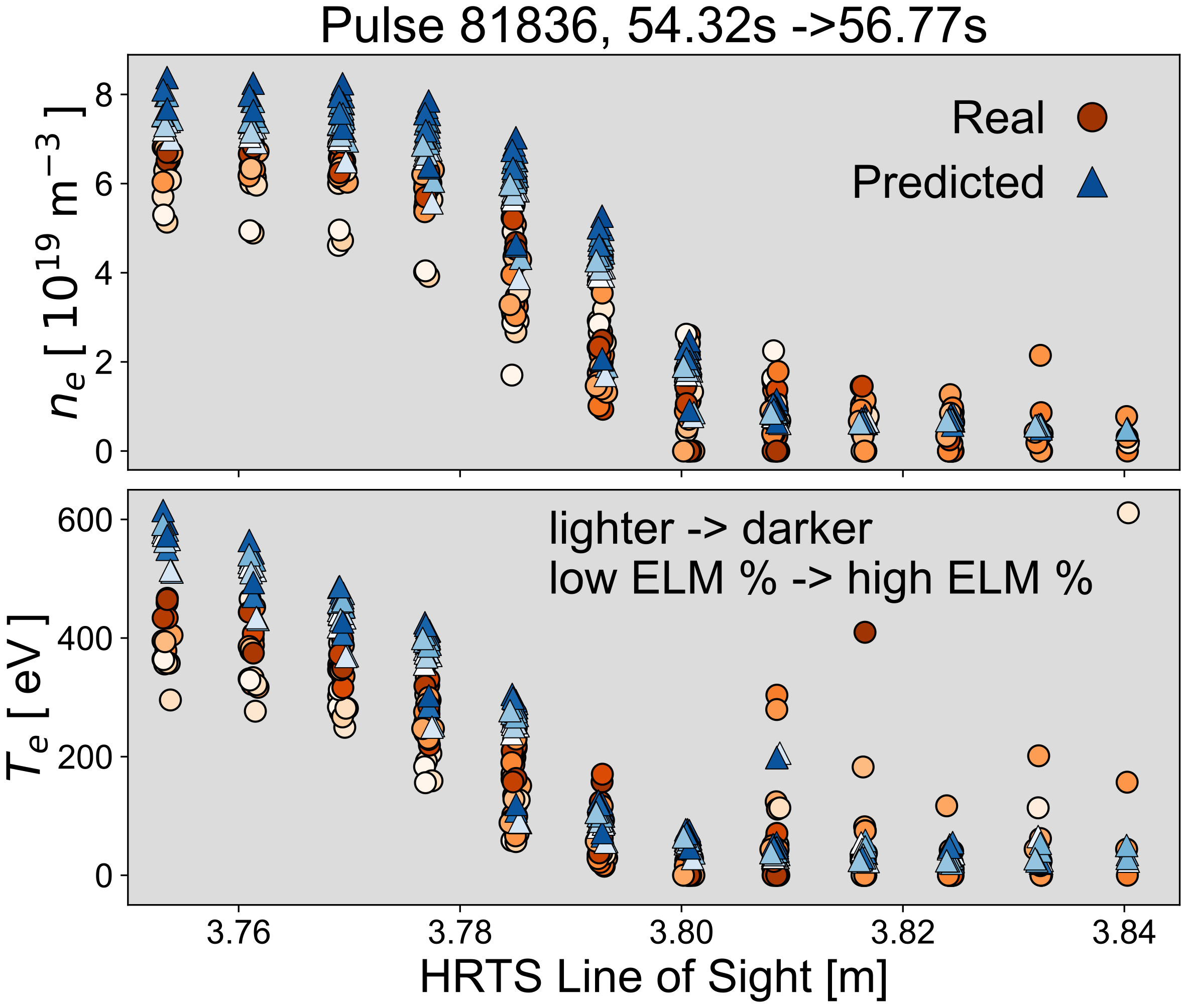}
    \end{minipage}
    \hfill
    \begin{minipage}[b]{0.45\textwidth}
    \includegraphics[width=\textwidth]{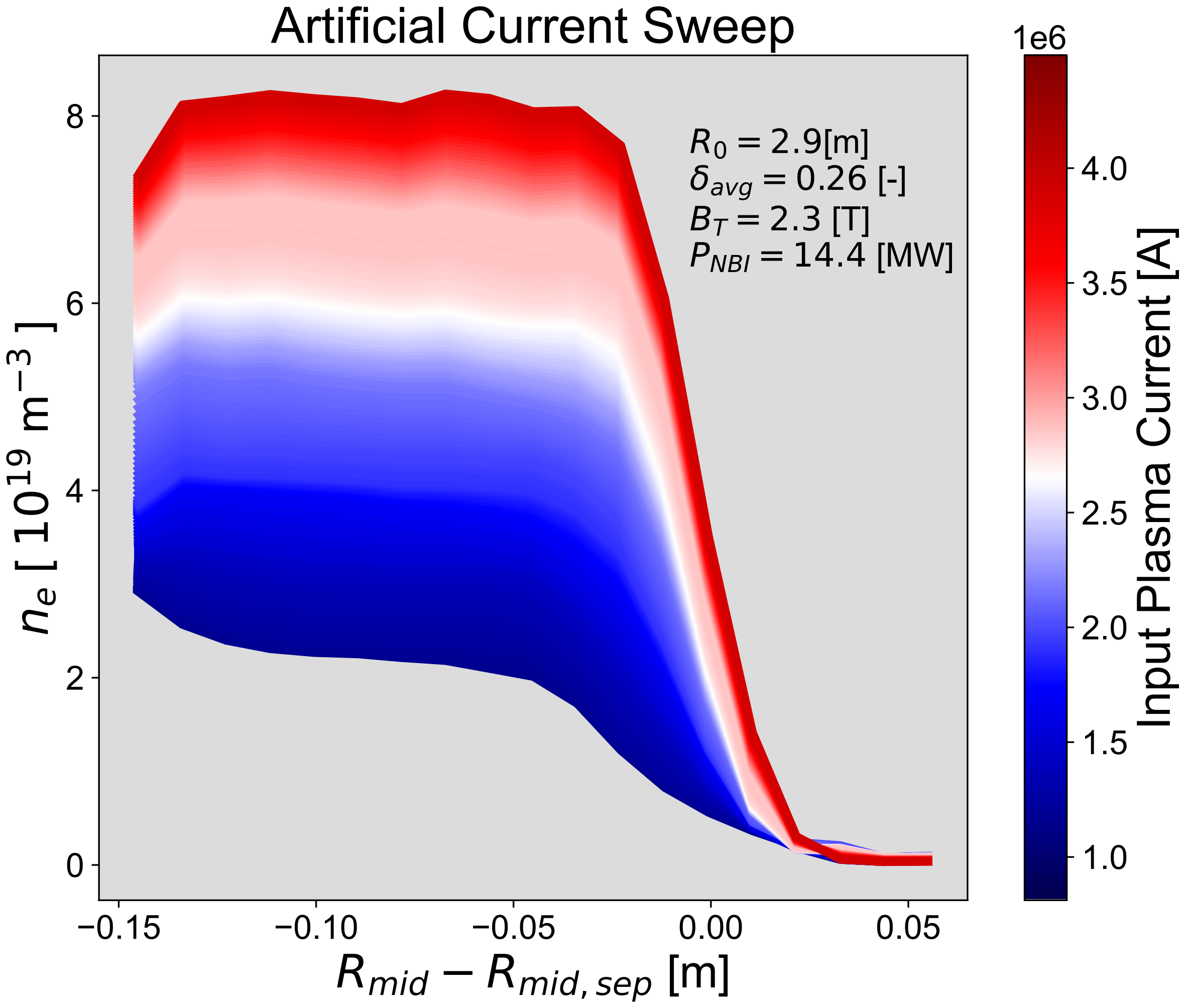}
    \end{minipage}
    \caption{One benefit of representation learning is the ability to physically inspect the latent representation as well as the output of the model. In both figures, the model conditionally generates density (and temperature) profiles given a set of machine parameters. \textbf{Right}: All machine parameters are kept constant, except for current, which is varied between 1 and 5 [MA]. The model captures the common proportionality between current and density, i.e., density positively linearly correlated with the plasma current.  \textbf{Left}: Machine parameters for pulse 81836 are passed to the model (previously unseen data), and the resulting profiles are plotted in comparison to the experimental HRTS measurements. The model learns the dependence of the ELM cycle, as the predicted density is lower for lower ELM \%s, seen by the lighter points.}
    \label{fig:current}
\end{figure*}

\begin{figure}
    \centering
    \includegraphics[width=0.45\textwidth]{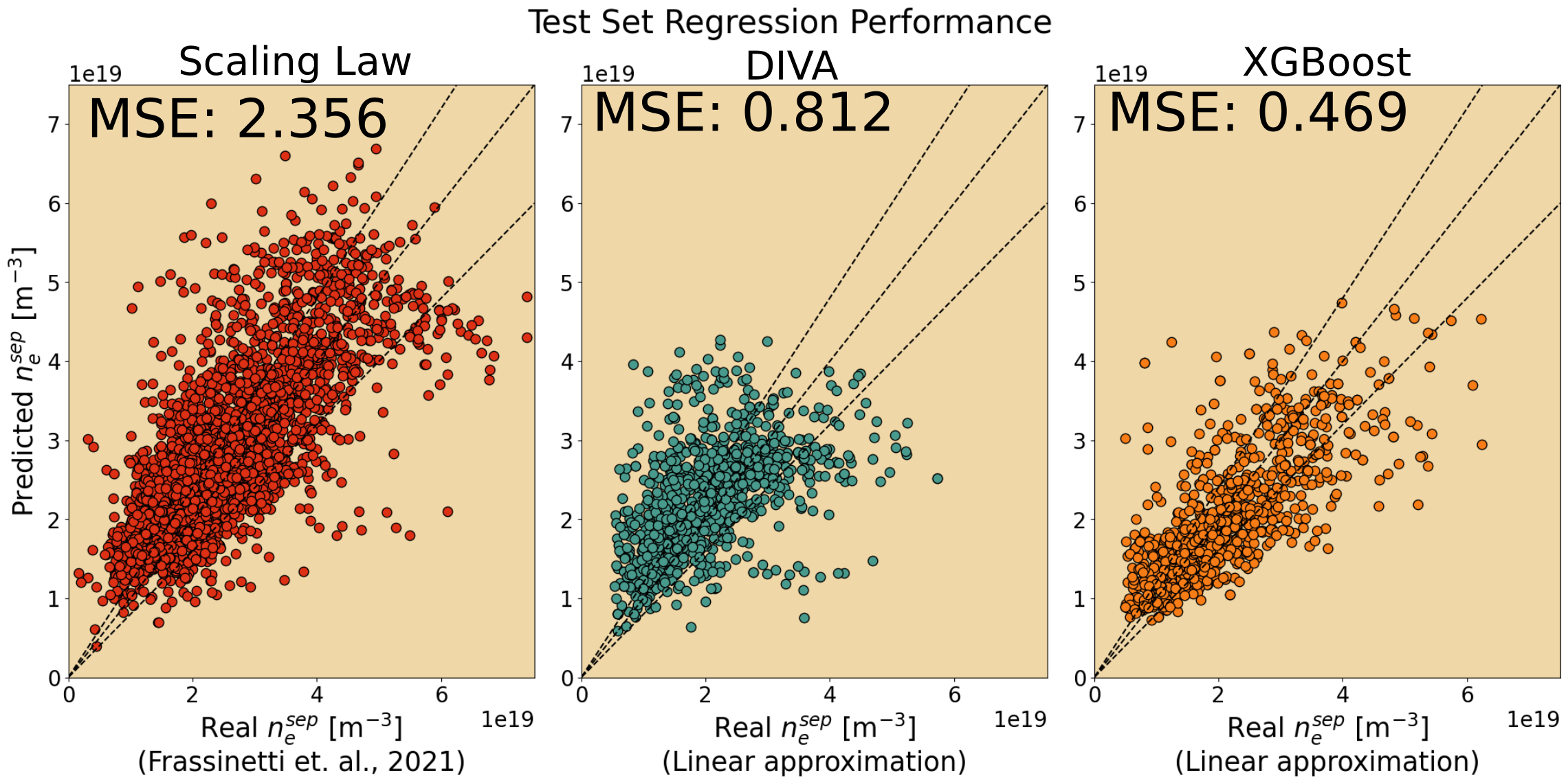}
    \caption{The results of predicting $n_{\text{e, sep}}$ with advanced machine learning techniques show that the direct mapping approach via XGBoost performs better than representation learning, while both outperform the log-linear scaling law from \cite{Frassinetti_NF_2021}. \textbf{Left}: Log-linear regression developed by Frassinetti et. al., \cite{Frassinetti_NF_2021} on the existing JET pedestal database.  \textbf{Middle}: The representation learning approach via DIVA framework is made on the created dataset, with predicted $n_{\text{e, sep}}$ estimated from conditionally generated profiles via the same linear approximation algorithm used in the dataset creation. \textbf{Right}: A direct mapping via XGBoost is made on the created dataset. Since the JET pedestal database has significantly less entries than the created dataset, entries are randomly sampled from the created dataset for a more reasonable qualitative comparison. In each figure, the solid black line represents unity, and the dashed black lines $\pm 20 \%$ error. The approximations of real $n_{\text{e, sep}}$ are noticeably larger in magnitude in the JET PDB compared to those in the created dataset. This is due to the lack of ELM averaging techniques for profile selection in this analysis, whereas the JET PDB approximations use only profiles from 75-99\% of the ELM cycle \cite{Frassinetti_NF_2021}.}
    \label{fig:RESULTS_ALL}
\end{figure}

For predicting $n_{\text{e, sep}}$, the direct mapping model currently outperforms the representation learning model (Fig. \ref{fig:RESULTS_ALL}). This is likely due to the added constrains for the latter. There, while predicting $n_{\text{e, sep}}$ we pass through latent variable $z$, which has a Gaussian prior and a relatively low dimensionality. This assumption about the data need not align with the true process that describes the plasma behavior: We can expect that this representation compresses away information about the exact location of $n_{\text{e, sep}}$. On the other hand, the direct mapping has no such limitation. Additionally, inferred separatrix location from the generated profile is very sensitive to the generated location $T_e \sim 100$ eV, and any error in this is propogated through the analysis. In the future, further constraints will be considered to better constrain the separatrix location in the generated profile. 

The direct mapping approach performs similar when fit using the JET pedestal database and the dataset created for this analysis (Fig. \ref{fig:RESULTS_ALL}). Note that the created dataset includes the fraction of the ELM cycle as a machine parameter, whereas the JET pedestal database does not. By expanding the database to include entries within the flat-top time window, the direct mapping can provide a decent approximation for inter-ELM $n_{\text{e, sep}}$ when given machine parameters and fraction of ELM cycle. Expanding existing databases and methods outside of the mtanh shows promise for training advanced machine learning methods. 

The representation learning model is observed to have learned the linear proportionality of plasma current and electron density, $I_P \propto n_e$ (Fig. \ref{fig:current}). The model is able to extrapolate this proportionality outside of the $I_P$ domain of the dataset (Table \ref{tab:domains}) to currents above and below 4 and 1 [MA] respectively. However, it would continue to increase the density with increasing current to arbitrarily low edge safety factors, whereas an instability would be observed in reality. This is likely because the data used for training does not explicitly contain the information about stability limits. However, these limits could be parameterized as part of the learning goal. Future studies will explore regularization of the latent space using parameterization for the boundaries of the stable operational domain.

Additionally, the model is observed to represent the cyclical crashing of the density over the course of an ELM cycle well (Fig. \ref{fig:current}). This shows that the learned representation of the plasma state potentially has the capability to generate full ELM-cycle consistent density profiles for a given set of machine parameters. However, this cannot decisively be said about the predicted temperature. 


\section{Conclusion}

Direct mapping and representation learning approach to predicting and inferring $n_\text{e,sep}$ have been explored. We evaluated both approaches on experimental data from JET, based on the EUROfusion pedestal database \cite{Frassinetti_NF_2021}. For solely predicting $n_\text{e,sep}$, the direct mapping performed significantly better than the more comprehensive representation learning approach. This indicates that for specific regressions task on a well labeled datasets, supervised, direct mapping machine learning algorithms can be very powerful.  However, the representation learning approach was able to learn a compressed representation of the pedestal electron density and temperature profiles, which could potentially have further applications in addition to predicting $n_\text{e,sep}$, e.g., control. Furthermore, future development of these algorithms and approaches are also expected to improve the prediction accuracy through inclusion of additional experimental information, e.g., temporal evolution of the plasma parameters. Additionally, we show how to propagate physics based constraints on deep learning models through approximations of the static pressure of the pedestal top, poloidal magnetic field and $\beta$. 
\section*{Acknowledgments}
This work has been carried out within the framework of the EUROfusion Consortium, funded by the European Union via the Euratom Research and Training Programme (Grant Agreement No 101052200 — EUROfusion). Views and opinions expressed are however those of the author(s) only and do not necessarily reflect those of the European Union or the European Commission. Neither the European Union nor the European Commission can be held responsible for them.


\begin{thebibliography}{9}
\bibitem{Frassinetti_NF_2021}
L. Frassinetti, \textit{et al.} Nucl. Fusion \textbf{61} (2021) 016001
\bibitem{XGBOOST}
T. Chen, \textit{et al.} KDD. \textbf{22} (2016) 2939785
\bibitem{VAE}
D. Kingma, \textit{et al.} ICLR. \textbf{2} (2014) 6114
\bibitem{DIVA}
M. Ilse, \textit{et al.} MIDL. \textbf{121} (2020) 
\bibitem{HRTS_system}
R. Pasqualotto, \textit{et al.} Rev. Sci. Instrum. \textbf{75} (2004) 3891
\bibitem{MTANH_Frass}
L. Frassinetti, \textit{et al.} Rev. Sci. Instrum. \textbf{83} (2012) 013506
\bibitem{MTANH_Groebner}
R.J. Groebner, \textit{et al.} Nucl. Fusion \textbf{41} (2001) 1789
\bibitem{Stangeby}
P.C. Stangeby, \textit{The Plasma Boundary of Magnetic Fusion Devices}(2000) 
\bibitem{Eich_2013}
T. Eich, \textit{et al.} Nucl. Fusion \textbf{53} (2013) 093031
\bibitem{Eich_NF_2018}
T. Eich, \textit{et al.} Nucl. Fusion \textbf{58} (2018) 034001
\bibitem{Kallenbach_2018}
A. Kallenbach, \textit{et al,} Plasma Phys. Control. Fusion \textbf{60} (2018) 045006
\bibitem{Stangeby_nesep}
P.C. Stangeby, \textit{et al.} Nucl. Fusion \textbf{55} (2015) 093014
\bibitem{Leonard_nesep}
A.W. Leonard,  \textit{et al.}  Nucl. Fusion \textbf{57} (2017) 086033
\bibitem{Dasbach_this_conference}
S. Daschbach, \textit{et al.} PSI-25 (2022), to be published...
\bibitem{TABOVER}
V. Borisov, \textit{et al.} CoRR \textbf{abs/2110.01889} (2021) \url{https://arxiv.org/abs/2110.01889}
\bibitem{Frass_NF_21_nesep}
L. Frassinetti \textit{et al.} Nucl. Fusion \textbf{61} (2021) 126054 
\bibitem{Wiesen_NM_2015}
S. Wiesen \textit{et al.} Journal of Nucl. Mater. \textbf{463} (2014) 480 )
\end{thebibliography}
\end{document}